\documentclass[prx,superscriptaddress,longbibliography,twocolumn]{revtex4-2}
\usepackage[final]{graphicx}% Include figure files
\usepackage{xfrac}
\usepackage{xcolor}
\usepackage{amsmath,amssymb}
\usepackage{float}
\usepackage[caption = false]{subfig}
\usepackage{braket}
\usepackage{hyperref}

\newcommand{\be}{\begin{equation}}
\newcommand{\ee}{\end{equation}}

\newcommand{\bea}{\begin{eqnarray}}
\newcommand{\eea}{\end{eqnarray}}
\newcommand{\beal}{\begin{align}}
\newcommand{\eeal}{\end{align}}

\begin{document}
%
%\preprint{}

\title{Characterizing spin-one Kitaev quantum spin liquids}

\author{Ilia Khait}
\affiliation{Department of Physics, University of Toronto, Toronto, Ontario, M5S 1A7, Canada}

\author{P.  Peter  Stavropoulos}
\affiliation{Department of Physics, University of Toronto, Toronto, Ontario, M5S 1A7, Canada}

\author{Hae-Young Kee}
\affiliation{Department of Physics, University of Toronto, Toronto, Ontario, M5S 1A7, Canada}
\affiliation{Canadian Institute for Advanced Research, Toronto, Ontario, M5G 1Z8, Canada}

\author{Yong Baek Kim}
\affiliation{Department of Physics, University of Toronto, Toronto, Ontario, M5S 1A7, Canada}
\affiliation{School of Physics, Korea Institute for Advanced Study, Seoul 02455, Korea}

\begin{abstract}
Material realizations of the bond-dependent Kitaev interactions with $S$=1/2 local moments 
have vitalized the research in quantum spin liquids. Recently, it has been proposed that higher-spin analogues of the Kitaev
interactions may also occur in a number of materials with strong spin-orbit coupling. 
In contrast to the celebrated $S$=1/2 Kitaev model 
on the honeycomb lattice, the higher-spin Kitaev models are not exactly solvable. Hence, the existence
of quantum spin liquids in these systems remains an outstanding question. In this work, we use
the density matrix renormalization group (DMRG) methods to numerically investigate the $S$=1 Kitaev model 
with both ferromagnetic (FM) and antiferromagnetic (AFM) interactions.
Using results on quasi-one-dimensional finite-size cylindrical geometries with circumferences of up to six legs,
we conclude that the ground state of the $S$=1 Kitaev model is a quantum spin liquid with a $\mathbb{Z}_2$ gauge
structure. We are also able to put an upper bound on the excitation gap.
The magnetic field responses for the FM and AFM models are similar
to those of the $S$=1/2 counterparts. In particular, in the AFM $S$=1 model, a gapless quantum spin liquid state emerges
in an intermediate window of magnetic field strength, before the system enters a trivial polarized state.
\end{abstract}

\maketitle

\section{Introduction}
A quantum spin liquid is a phase of matter characterized by long-range entanglement and fractionalized excitations in magnetic systems described by spin models~\cite{Savary_2016,Trebst_2017,Zhou_2017,Knolle_2019,Broholmeaay0668}. While fascinating, its existence has been a subject of long debate until the exactly solvable Kitaev model was found~\cite{Kitaev_2006}. The $S$=1/2 Kitaev model on a honeycomb lattice is described by bond-dependent Ising interactions which lead to strong frustration. The excitations about the ground state of the $S$=1/2 Kitaev model are visons ($\mathbb{Z}_2$ fluxes) and Majorana fermions. When the time-reveral symmetry is broken, for example by a magnetic field, this phase becomes a chiral spin liquid with gapless Majorana fermions propagating along the boundary of the system, leading to the half-quantized thermal Hall conductivity~\cite{Kitaev_2006}. Recently a microscopic derivation on how to realize the Kitaev model in solid-state materials has been established, where strong spin-orbit coupling in strongly correlated Mott insulator is an essential ingredient~\cite{Jackeli_2009}. Since then candidate materials such as honeycomb iridates~\cite{Singh_2012,Krempa_2014,Rau_2016,Winter_2017} and RuCl$_3$~\cite{Plumb_2014,Winter_2017} have been proposed. Strikingly, the half-quantized thermal Hall conductivity under a magnetic field in RuCl3 was recently reported~\cite{Kasahara_2018}. While further experimental evidences are required, RuCl$_3$ seems to offer a playground to study exotic physics in correlated systems with spin-orbit coupling.
	
In parallel, there have been questions on whether higher spin Kiatev models may possess anyonic excitations similar to $S$=1/2 model. The higher spin Kitaev model is no longer exactly solvable, even though one can find a plaquette operator that commutes with the Hamiltonian, i.e, there is a conserved quantity. This model has been of a theoretical interest until a microscopic route to higher spin model was found~\cite{Peter_2019}. The Hund’s coupling of the transition metal ions together with the spin-orbit coupling of anions generate the bond-dependent Ising interactions. An example of $S$=1 candidate materials was also proposed~\cite{Peter_2019}. Despite such progress, due to the fact that the higher-spin Kitaev models are not exactly solvable, the nature of quantum ground states and the transition to the polarized state under the magnetic field remain outstanding subjects of theoretical investigation~\cite{Koga_ED_2018,Oitmaa_2018}.

\begin{figure}
	\includegraphics[width=\linewidth]{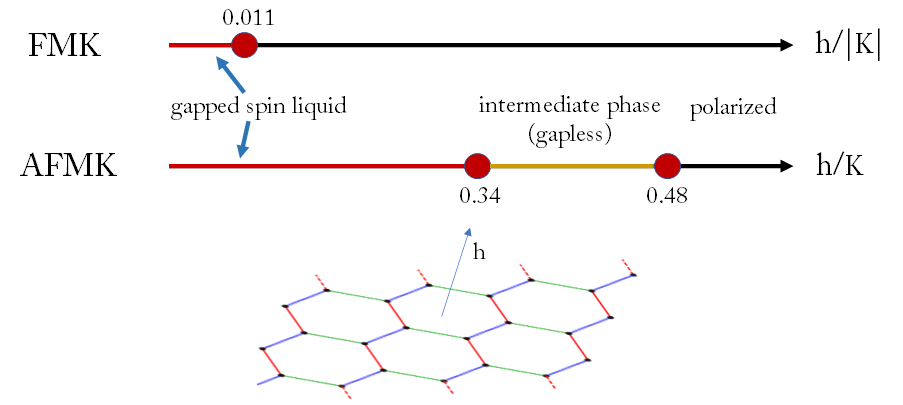}
	\caption{Illustrative phase diagram of the $S$=1 Kitaev model with ferromagnetic (FM) and antiferromagnetic (AFM) couplings as a function of applied magnetic field in the $[111]$ direction. For the AFM Kitaev case, we find numeric signatures for two phase transitions between a spin-liquid at zero field ($h=0$) and a polarized phase at strong fields. For FM Kitaev, we only find a single transition at a relatively weak field. The different phases are identified using the entanglement spectrum and the apperant gap closing between the ground state and the first excited state (see text for further elaborations). }
	\label{fig:PD}
\end{figure}
	
Here we study the $S$=1 Kitaev model and address the questions raised above using DMRG. Some basic properties of the $S$=1 Kitaev model on the honeycomb lattice are known~\cite{Baskaran_2008}. In analogy to the $S$=1/2 model, one can define a plaquette flux operator $W_p$ on each hexagon, which commutes with the Hamiltonian. 
These plaquette operators are written in terms of the $\pi$-spin-rotation unitary operators for an arbitrary spin-$S$ quantum number~\cite{Baskaran_2008}.
Hence, each eigenstate can be labelled by the eigenvalues of these flux operators. This constant of motion can be used to show that spin-spin correlations exist only for nearest neighbours, however this by itself is not enough to allow for exact solutions.
Exact diagonalization (ED) studies up to $24$-site clusters concluded that the ground state might be a gapless quantum spin liquid.
On the other hand, a recent tensor network construction of the variational wavefunction suggests that the ground state is a gapped $\mathbb{Z}_2$ spin liquid with abelian quasiparticles. 
Since each numerical approach~\cite{Tensor_2019} has its own limitations, it is important to synthesize the efforts from different numerical and analytical approaches for the ultimate understanding.

In this work, we study the $S$=1 Kitaev model on the honeycomb lattice using DMRG~\cite{White_DMRG_1992,Schollwock_2011,ITensor} on a cylindrical geometry with various circumference lengths. We start with two-leg ladder systems (or $L_y=2$), where we study system sizes up to $250$ sites ($L_x=125$). It is worthwhile to note that the ladder geometry offers a valuable insight to two-dimensional system despite its obvious limitation. In the case of the $S$=1/2 Kitaev model, the Kitaev spin liquid has even-odd effects depending on the number of legs. While the phase is gapped for the two-leg ladder, it still exhibits Majorana fermions as a zero energy state with open boundary conditions~\cite{DeGottardi_2011, Hur_2017}. Furthermore, in the extended Kitaev model, i.e. the Kitaev-Heisenberg model, not only the phases but also the transition between the phases were captured in the ladder model~\cite{Feng_2007,Catuneanu_2019}. With an additional magnetic field, the intermediate phase characterized by a staggered chirality has been identified~\cite{Sorensen_2020}. This is similar to the intermediate phase found in spin-1/2 Kitaev model with a honeycomb geometry~\cite{Zhu_2018,Hickey_2019,Kaib_2019}. 

The ground state has a uniform flux, for which $W_p = 1$ for any hexagonal plaquette. We demonstrate that the spin-flip operator at a given site generates two adjacent ``vortex" plaquettes with $W_p = -1$, just like in the $S$=1/2 case. From the two-fold degeneracy of entanglement spectrum (ES), we conclude that the ground state of the two-leg ladder system is a symmetry-protected topological (SPT) phase, with a two-fold degenerate ground state. This result is similar to the two-leg ladder system of the $S$=1/2 model~\cite{Catuneanu_2019}, albeit the degeneracy structure of the ES in the $S$=1 model is different from that of the $S$=1/2 model.

Given that the $S=1$ model naturally offers an AFM Kitaev exchange interaction~\cite{Peter_2019} unlike the $J_{\rm eff}=1/2$ FM Kitaev model, it is worthwhile to investigate the phase diagram under the magnetic field. We apply fields perpendicular to the honeycomb plane (parallel to the [111] direction). 
In the $S$=1/2 model~\cite{Nasu_2018,Janssen2019}, it is known that the magnetic field phase diagram of
the two-leg ladder system is very similar to that of clusters with wider circumferences~\cite{Feng_2007}. Examining the magnetic field responses of the FM and AFM Kitaev couplings, we find that the magnetic field dependence is surprisingly similar to the $S$=1/2 case~\footnote{See Appendix B for an overview}.
For example, an intermediate phase exists for the AFM model 
in a window of magnetic field strengths, right before the system enters a polarized state, while for the FM model there is a direct transition to the polarized state at a much lower critical field.
On the other hand, in comparison to the $S$=1/2 model phase diagram, the zero-field ground state of the AFM model is much more robust against the magnetic field than that of the FM case.

\begin{figure}
	\includegraphics[width=\linewidth]{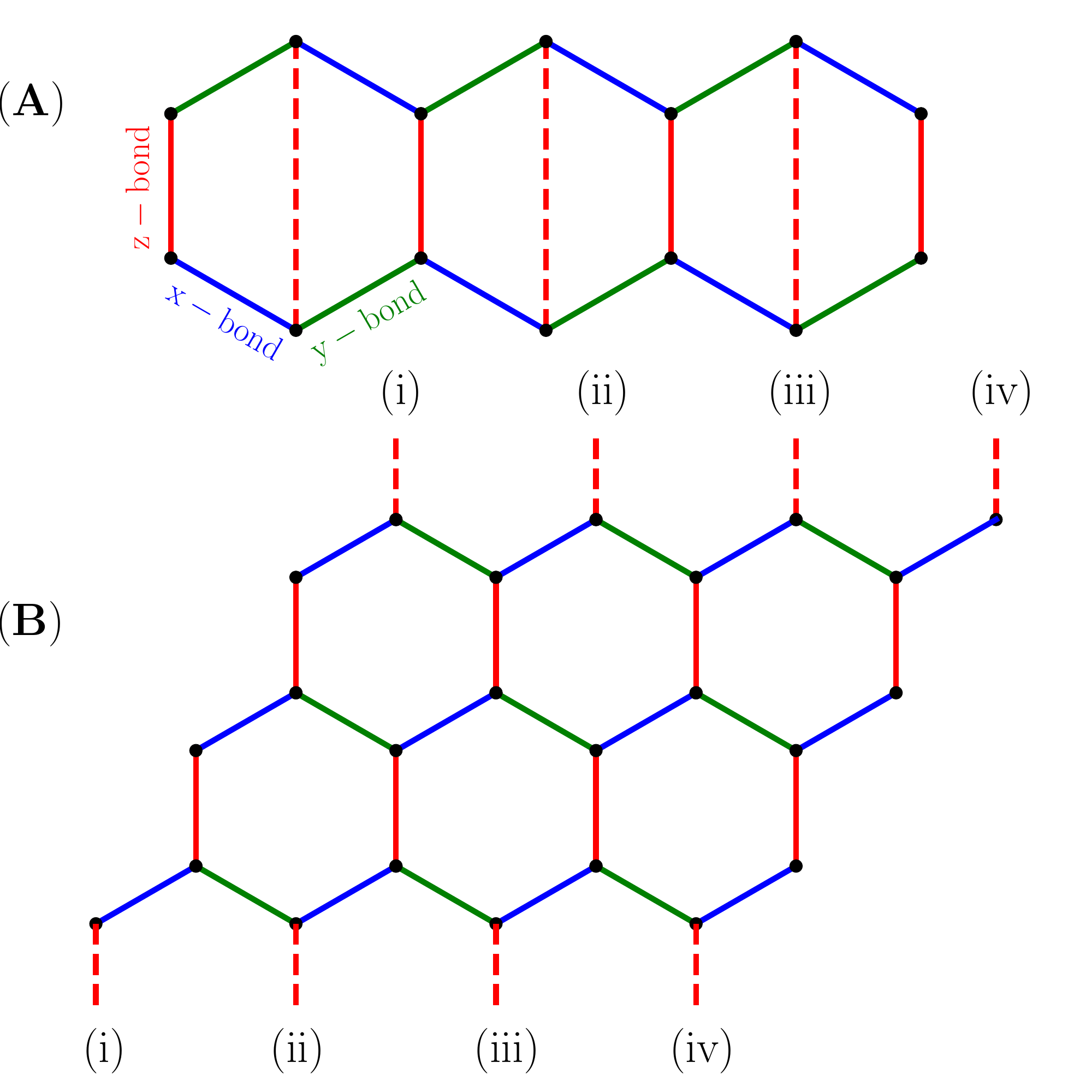}
	\caption{Different geometries for the honeycomb Kitaev model. Blue, green, and red lines represent the x-bonds, y-bonds and z-bonds respectively. Panel A is the two-leg ladder ($L_y=2$), which is equivalent to a square ladder. Panel B shows the three-leg ladder ($L_y=3$) with periodic boundary conditions along one of the axes. Red dashed lines represent the periodic links, and roman letters represent the periodic z-bonds.}
	\label{fig:Geom}
\end{figure}

We then consider a three-leg (or $L_y=3$) system with periodic boundary conditions along the circumference, with cluster sizes of up to $144$ sites ($L_x=48$). 
The ES no longer shows any degeneracy. Furthermore, the ground state energy of the FM and AFM models is exactly the same, just like for the $S$=1/2 model. We define the Wilson loop operator along the circumference direction $W_{\ell}$,
which commutes with the Hamiltonian. 
We further investigate the ground state of the AFM model, and its two lowest energy excited states. The first three lowest energy states are in the $W_\ell = + 1$ sector. The lowest excitation energy  (the energetic difference between the first excited state energy and the ground state), for the largest system at hand with 72 sites, $L_x = 24$, is $\Delta = 5 \times 10^{-2} ~\rm K$, where $K$ is strength of the Kitaev interaction. This is our {\it upper bound} on the excitation gap of the $S$=1 AFM Kitaev model. 
We also study how the ground state topological properties change with increasing the number of legs. We use systems of up to $48$ sites with $L_y \leq 6$, and find that the ground states of $L_y=4~{\rm and }~6$ clusters are in the $W_{\ell} = -1$ sector, while for $L_y =3~{\rm and}~ 5$ clusters
it is in the $W_{\ell} = +1$ sector . Therefore it appears that there is an even-odd effect, which suggests that the ground states
in $W_{\ell} = +1$ and $W_{\ell} = -1$ sectors may become degenerate
in the thermodynamic limit. This would be consistent with the 
two degenerate ground states on the cylinder for $\mathbb{Z}_2$ spin liquid states.

The effect of external magnetic fields is studied for the three-leg cylinder consisting of 24-sites. The overall behaviour is similar to the results of the two-leg ladder system. That is, the critical field for the AFM model is much larger than that of the FM case. Due to slow convergence, however, we can only see the first transition to the intermediate state in the AFM model for the $L_y=3$ system. The transition is also well apparent in other observables such as the entanglement entropy (EE). We investigate the excitation energy gap as a function of an external magnetic field and find that it vanishes as one approaches the aforementioned transition. This is consistent with
the picture of a gapless intermediate phase, just like for the $S$=1/2 AFM model subject to magnetic field~\cite{Zhu_2018,Hickey_2019,Patel_2019}.
The general overall picture emerging from these studies is that
the ground state of the $S$=1 Kitaev model is a quantum spin liquid with a
$\mathbb{Z}_2$ gauge structure, and that the response to external magnetic fields is very similar to the $S$=1/2 case. 
Our current numerical data may be consistent with a quantum spin liquid with a small excitation gap. This, however, does not preclude a gapless spin liquid in the thermodynamic limit. 

The rest of the paper is organized as follows.
In Section II, we introduce the model and explain basic symmetry properties.
Here we also discuss the cylindrical geometry we use throughout this manuscript, and briefly explain
the DMRG calculation details. We present the results of the two-leg ladder systems,
or $L_y=2$, and  the  three-leg system, $L_y = 3$, in Section III. Here we also briefly discuss more results
on systems with up to six-leg ladders ($L_y = 6$). 
In Section IV, we discuss the implications of our results.

\section{Model}
The Kitaev model Hamiltonian is given by
\be
\mathcal{H} =  K \sum_{\substack{\gamma \\ \left\langle i,j \right\rangle_\gamma}} S_i^\gamma S_j^\gamma - \vec{h} \cdot \sum_i \vec{S}_i,  
\label{eq:Ham1}\ee
where $S_i^\gamma$ is the $\gamma$ component of an $S=1$ spin at site $i$ on a honeycomb lattice, $\left\langle i,j \right\rangle_\gamma$ is two nearest-neighbour sites along an $\gamma$ bond ($\gamma=x,y,z$). Throughout this manuscript we focus on the isotropic Kitaev model. A natural extension would be to consider bond-dependent couplings, i.e.  $K_\gamma \neq K_{\gamma'}$. This was recently studied in Ref.~\cite{Lee_2020}, where it was found that the spin liquid is stable against small anisotropy in the exchange couplings.  $\vec{h}$ is a uniform magnetic field, where we discuss a field along the $[111]$ direction. Applying this field breaks time-reversal symmetry. For the $S$=1/2 model, this opens a gap in the fermionic spectrum, and in the perturbation theory~\cite{Kitaev_2006}, a three spin interaction $S^x_i S^y_j S^z_k$ is generated.

\subsection{Symmetries}
It  was shown that the pure ($h=0$) Kitaev model on a honeycomb lattice, which is defined in Eq.~\ref{eq:Ham1}, has a constant of motion defined on a plaquette~\cite{Baskaran_2008}
\be
W_p^j = \prod_{i \in \mathcal{P}^j} e^{i\pi S_i^\alpha},
\label{eq:Wp}\ee
where $\mathcal{P}^j$ is the $j$-th plaquette consisting of six sites on a single hexagon, and $\alpha$ represents the protruding bond along $\mathcal{P}$. 
\begin{figure*}
	\includegraphics[width=\textwidth]{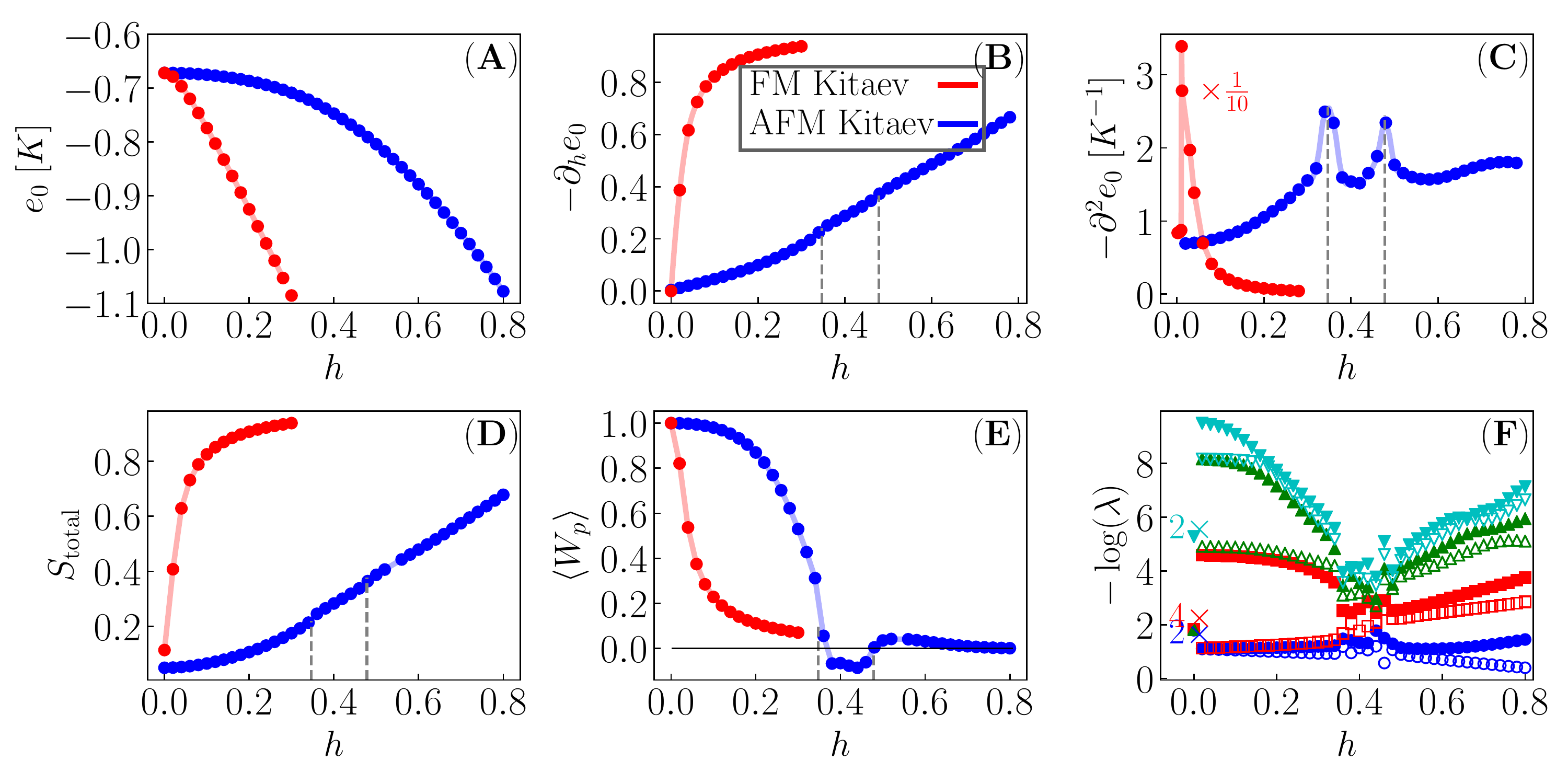}
	\caption{Magnetic phase diagram of the ground state of the $S=1$ Kitaev model on a two-leg ladder, as a function of a uniform magnetic field, $h$ in units of the Kitaev interaction $K$, in the $[111]$ direction. Panel A shows the ground state energy. Panel B is the magnetization density. Panel C is the uniform magnetic susceptibility, which clearly shows a transition to the polarized state at $h_c^{\rm FM} = 0.011$ for FM,
	and $h_{c_1} \approx 0.34$ and $h_{c_2} \approx 0.48$ for the AFM model. Panel D is the magnitude of the total spin. Panel E is the plaquette operator's, $W_p$, expectation value. Panel F is the entanglement spectrum (ES), for the AFM Kitaev partitioned with a cut on the middle rung. At $h=0$, the ES has a $2-4-2$ degeneracy structure, which is depicted by the numbers to the left of the markers. As can be seen in this figure, the degeneracy breaks as finite magnetic field is introduced. Note that observables B-D and F show clear signatures of two phase transitions. Dashed lines represent the critical fields $h_{c_1}$ and $h_{c_2}$ marking the intermediate phase.}
	\label{fig:MagPD}
\end{figure*}
We define the Wilson loop operator along the circumference of the cylinder, which commutes with the Hamiltonian
\be
W_\ell = \prod_{ i \in \rm y-loop} e^{i\pi S_i^y},
\label{eq:Wl}\ee
where $y$-loop is a closed loop around the circumference in the periodic direction. 

For the $S=1/2$ Kitaev model, it can be shown that spin operators $S_i^\alpha$ acting on any eigenstate of the Hamiltonian, would lead to a $\pi$-flux insertion, or a sign-flip of the associated plaquette operator $W_p^j$, where $j$ corresponds to two adjacent plaquettes which include site $i$, and share an $\alpha$-bond. Since $W_p^j$ commutes with the Hamiltonian, it makes the spin-spin correlation function short-ranged, such that it is non-zero for nearest neighbours and exactly vanishes for further neighbours, $\left\langle S_i^\alpha S_j^\beta \right\rangle \propto \delta_{\alpha\beta} \delta_{\left\langle i,j \right\rangle _\alpha}$. This property remains the same for the $S=1$ Kitaev model (as was previously semi-classically proven for higher spin~\cite{Baskaran_2008}).

\subsection{Geometries}
In this study we mainly focus on two geometries, one of a two-leg ladder and one of a three-leg ladder (see Fig.~\ref{fig:Geom}). It was previously shown~\cite{Feng_2007} that the two-leg ladder's restricted geometry is able to capture the phase transitions for the $S$=1/2 model fairly well. Later in~\cite{Catuneanu_2019}, it was demonstrated that this simplified geometry's phase diagram in the Kitaev-Heisenberg plane is very similar to the 2D phase diagram on the honeycomb lattice. Hence, we expect that any transition to be found for the $S=1$ ladder would also appear in the 2D phase diagram albeit with different critical parameters (field strength, etc.). An additional advantage is that it is much easier, numerically, to access very large systems due to shorter-range interactions.  

The three-leg ladder is the minimal geometry which allows the Wilson loop operator along the circumference $W_\ell$. In addition, as discussed in Ref.~\cite{Gohlke_2018}, it allows probing the high-symmetry $K$-points in the Brillouin zone, which host Dirac fermions in the spin-half case.

\section{Results}
We study the spin-one Kitaev model on a honeycomb lattice, using the DMRG technique. Similarly to numerous previous studies, this method~\cite{White_DMRG_1992,Schollwock_2011,ITensor} can be used to infer useful information about the 2D limit by utilizing quasi-one-dimensional geometries and finite-size scaling.
We use cylinders with open boundaries conditions, with up to $250$ sites for the two-leg geometry, or $L_x=125,\ L_y=2$, (Fig.~\ref{fig:Geom} A), up to $144$ sites for the three-leg geometry, or $L_x=48,\ L_y=3$, (Fig.~\ref{fig:Geom} B). We retain 7200 states in the reduced density matrix, with no symmetries kept, and we found that 45 sweeps were sufficient for good convergence.  The DMRG relative truncation error was less than $10^{-9}$. 

\begin{figure}
	\includegraphics[width=\linewidth]{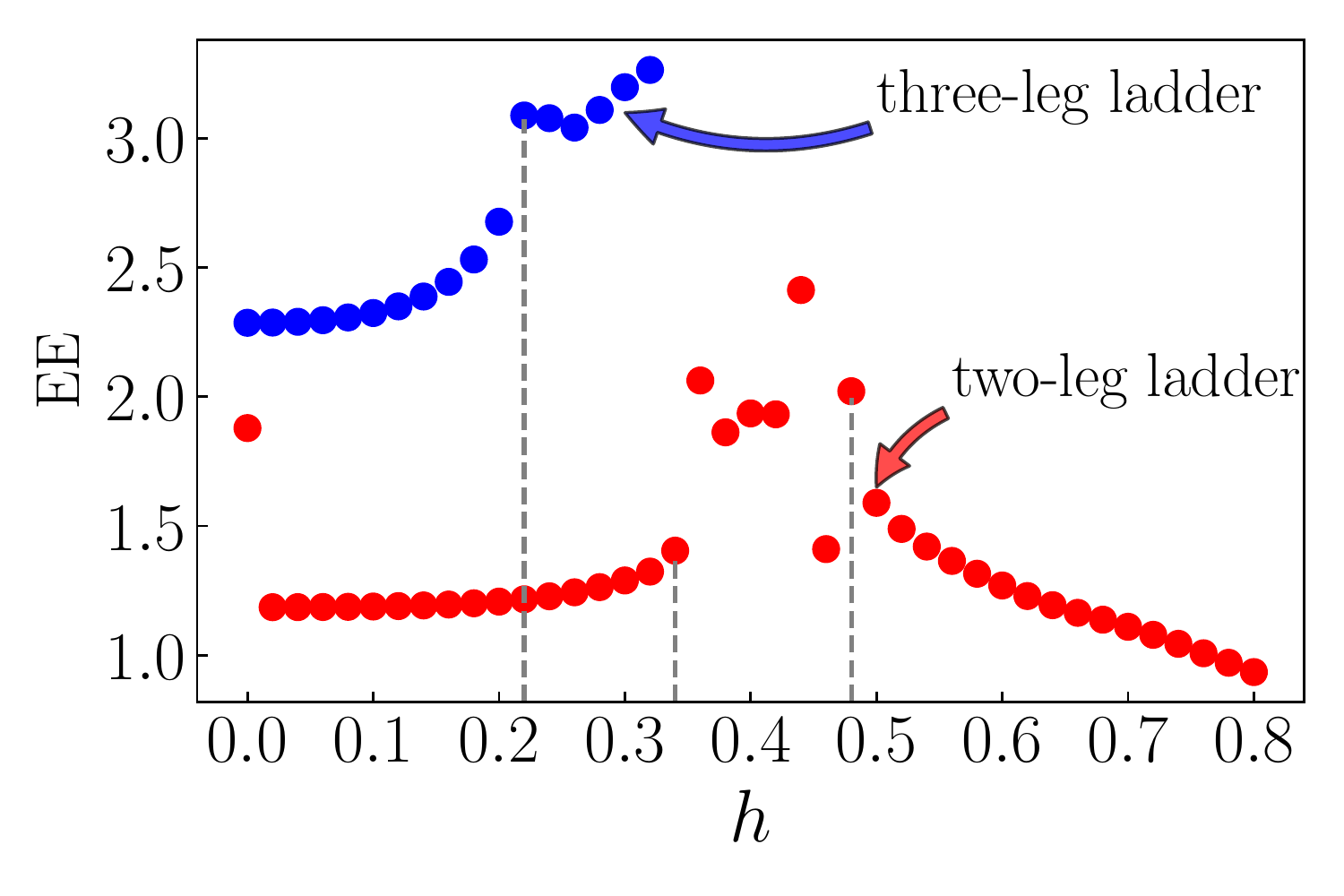}
	\caption{Entanglement entropy (EE) as a function of magnetic field $h$. Red circles depict the two-leg ladder ($250$ sites), and blue circles depict the three-leg ladder ($24$ sites with cylindrical boundary conditions). A jump at $h=0$ is seen for the two-leg geometry, which indicates the breakdown of the symmetry protected topological (SPT) phase due to the applied field. At higher fields the entropy increases, until it starts to wiggle due to the phase transition to the intermediate phase for $h_{c_1}<h<h_{c_2}$ (grey dashed lines, exact values are in the main text). Similar behaviour is seen for the three-leg ladder, where the transition to the intermediate phase is captured by the EE. For the entanglement spectrum see Fig.~\ref{fig:MagPD} F and Fig.~\ref{fig:ES_Ny_3}.}
	\label{fig:EE}
\end{figure}

%\subsection{General ground state properties}
We consider both the FM and AFM Kitaev couplings ($K= \mp 1$ in Eq.~\ref{eq:Ham1}). Note that the AFM couplings were shown to host a wider Kitaev spin liquid phase in the parameter space of the Kitaev-Heisenberg model~\cite{Peter_2019}. Moreover, one should note that under addition of a sufficiently small nearest-neighbour Heisenberg interaction, the following results are still valid, and that the phase diagrams presented here remain qualitatively the same. 

{\it Wilson loop operators: $W_\ell$ - }
We examine ladders of up to $L_y=6$. One can define the $W_\ell$ operator (see Eq.~\ref{eq:Wl}) on a geometry consisting of three legs and above. We find an even-odd effect for different topological sectors. For the odd-numbered leg systems, ($L_y=3,5$), we find the ground state to be in the $W_\ell=1$ sector, while for the even-numbered leg clusters, ($L_y=4,6$), the ground state is in the $W_\ell=-1$ sector.   

{\it Short range correlation effects - }
The spin $S=1$ model hosts very short spin-spin correlations. For the pure Kitaev limit, $h=0$ point, spin-spin correlations are non-zero only along the specific bonds, i.e. $\left\langle S_i^\alpha S_j^\beta \right\rangle \propto \delta_{\alpha\beta} \delta_{\left\langle i,j \right\rangle _\alpha}$. This is a direct consequence of the role of spin operators acting on eigenstates of the Hamiltonian - they thread $\pi$-flux into two adjacent plaquettes, i.e. they change the local plaquette expectation value by a factor of $-1$. 
However unlike the $S=1/2$ case, where the flux insertion operators are Pauli matrices, the spin $S=1$ operators alter the normalization of the state. Hence the resulting state after the spin flip should be written as the following state $\ket{\psi}$, where two of the adjacent plaquettes to site $i$, sharing an $x$-bond, gain the additional $\pi$-flux: $\ket{\psi} = S_i^x\ket{{\rm GS}} \left(\braket{{\rm GS}|(S_i^x)^2|\rm {GS}}\right)^{-1/2}$, where $\ket{\rm {GS}}$ is the ground state (this property actually applies to any eigenstate). Moreover, at finite field strengths, $h$, although $W_p$ is not a conserved quantity anymore, applying the spin operators still flips the sign of the two adjacent plaquttes.
Furthermore, $W_p$ does not commute with the Hamiltonian, and longer-ranged correlations appear and the property that only bond-dependent correlations are present and non-zero is lost, i.e. $\left\langle S_i^\alpha S_j^\alpha \right\rangle \neq 0$ for any two sites $(i,j)$.

\subsection{Two-leg ladder}
\begin{figure}
	\includegraphics[width=\linewidth]{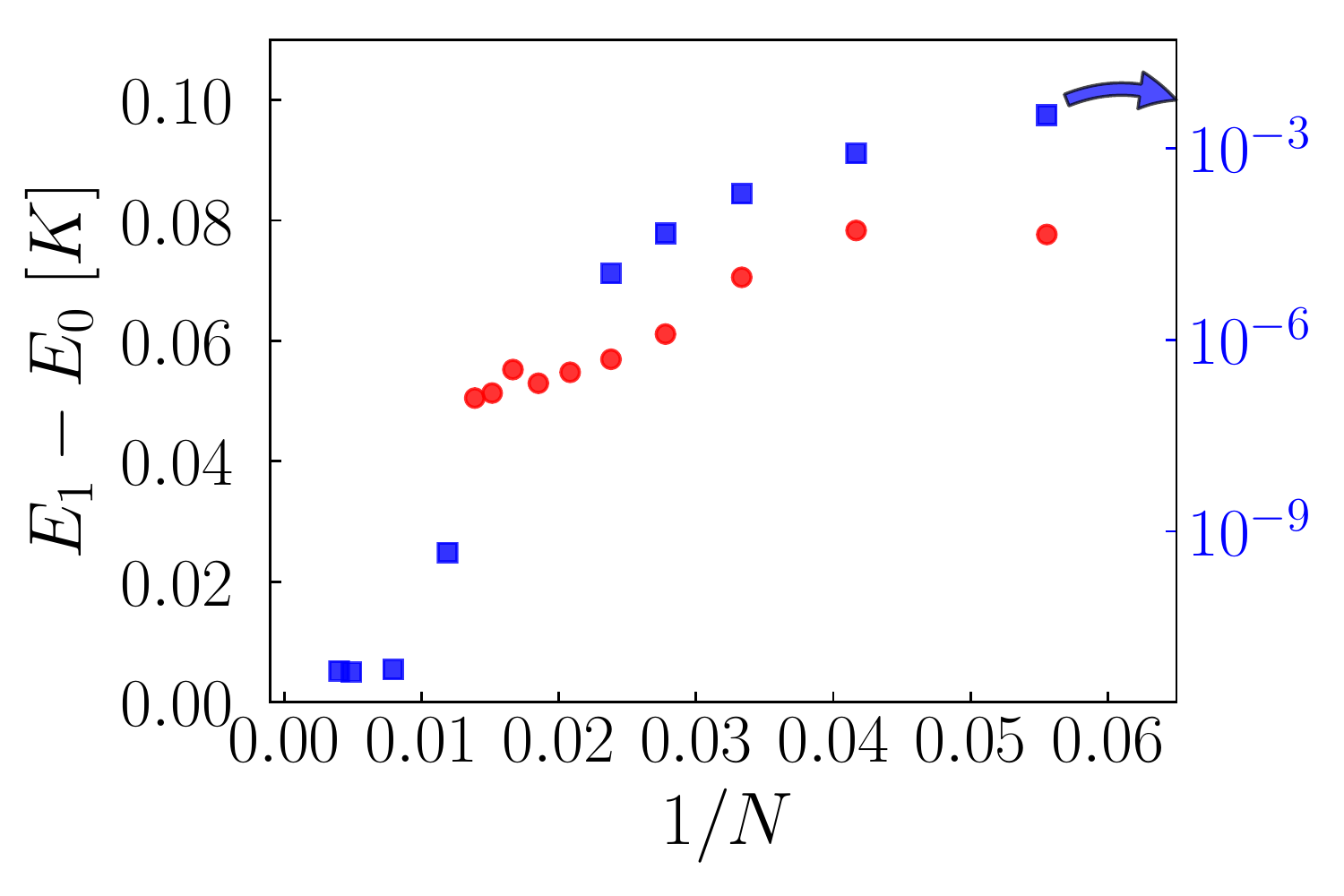}
	\caption{ Excitation gap of the AFM Kitaev model on two-leg ladder (blue) and three-leg ladders (red) as a function of inverse cluster size $N$. For the three-leg ladder a naive extrapolation to the thermodynamic limit would suggest that the ground state of the AFM model is gapped, while for the two-leg ladder it is an SPT with a degenerate ground state. Note that for the three-leg ladder both the ground state and the first excited state are in the $W_\ell = 1$ sector.}
	\label{fig:gapNy_3}
\end{figure}

%\subsection{Ground state properties} 
The two-leg ladder, with boundary conditions depicted in Fig.~\ref{fig:Geom} A, exhibits a doubly degenerate ground state in the thermodynamic limit (the energy difference between the ground state and the first excited state with $L_x=125$ is $\approx 2 \cdot 10^{-12}$, see Fig.~\ref{fig:gapNy_3})~\footnote{Even for relatively small systems with $L_x=8$ the two states are degenerated.}. For both FM and AFM coupling the ground state energies are the same, as shown in Fig.~\ref{fig:MagPD} A, and the two degenerate states have uniform flux, $W_p = 1$. Interestingly, for the AFM coupling, the degeneracy is present for $h < 0.2$, and the energy difference between the two lowest energy states is $\approx 10^{-12}$. At $h=0.2$ the gap density is $\approx 3\cdot10^{-4}$, which can still be attributed to finite size effects.

%\subsection{Magnetic phase diagram}
The magnetic field dependence of the two-leg ladder, is summarized in Fig.~\ref{fig:MagPD}. For both FM and AFM couplings, we plot the ground state energy, the magnetization density, and the uniform magnetic susceptibility, as a function of magnetic field parallel to the $[111]$ direction (panels A-C respectively of Fig.~\ref{fig:MagPD}). 
For the FM there appears to be a single transition, at weak fields, to a polarized phase, which is manifested in our data as a divergence at low fields in the magnetic susceptibility (Fig.~\ref{fig:MagPD} C). The critical field is $h^{\rm FM}_c = 0.011$.  However, for the AFM coupling, one notices two kinks around $h_{c_1} \approx 0.34$ and $h_{c_2} \approx 0.48$ in the magnetization, and two peaks at the same positions for the magnetic susceptibility. 
While the magnetization density parallel to $[111]$ is increasing as expected, the perpendicular (in plane) magnetization remains negligible regardless of the strength of the magnetic field applied. The spin size, $\left| S_{\rm total} \right|$ (Fig.~\ref{fig:MagPD} D) is extracted from the ground state expectation value $\left\langle  S_{\rm total}^2 \right\rangle$. It displays a two-kink structure at the same critical fields.   
Unlike the two previous observables, from the plaquette operator, $W_p$, whose expectation value is given in Fig.~\ref{fig:MagPD} E, one cannot identify the exact location of the phase transitions. Moreover, since in the presence of magnetic field it is no longer a constant of motion, the decrease in its value goes hand in hand with the increase of longer-ranged spin-spin correlations. 

The two peaks in the magnetic susceptibility for the AFM Kitaev (Fig.~\ref{fig:MagPD} C) indicate three phases the phase diagram. For $h<h_{c_1}$, the magnetization starts to build up, and the susceptibility and total spin are increasing non-monotonically. By solving the entire spectrum for small clusters using ED, we see that the distance between eigenenergies begin to shrink (as shown in Fig.~\ref{fig:ED_18}), i.e. the density of states at low energies is increasing, and, finite size-gaps begin to decrease. Once the spectrum collapses at $h_{c_1}$, a new phase appears for fields $h_{c_1}<h<h_{c_2}$. This phase is characterized by the peaks appearing in the magnetic susceptibility. This intermediate phase, which shows the least convergence, and requires much larger bond-dimensions ($>7200$ kept states), could possibly have a diverging correlation length, and hence hinting that it might be a gapless phase - consistent with the spectrum collapse. Finally, a polarized phase appears for fields $h>h_{c_2}$, where a linear increase in the magnetization and the total spin is seen.

{\it Entanglement spectrum and entanglement entropy - }
An SPT phase can be characterized by double degeneracy of its ES~\cite{Pollmann_2012}, and, indeed, the full ES is degenerate here. In fact, at $h=0$ the degeneracy pattern of the density matrix eigenvalues, $\lambda_i$ is $2-4-2$ for the entire ES (see Fig.~\ref{fig:MagPD} F). This aforementioned degeneracy is broken in the presence of magnetic field. In addition, the two transitions to the intermediate phase, and to the polarized phase are revealed by examining the EE as shown in Fig.~\ref{fig:EE}. At the pure Kitaev limit, $h=0$, the entropy is high, it then jumps to a lower value in the presence of a weak magnetic field, indicating a phase transition from the Kitaev limit. As one further increases the field, the EE rises until it jumps once more at $h=h_{c_1}$, indicating the intermediate phase, then there exists another jump at $h=h_{c_2}$. After increasing the field beyond $h=h_{c_2}$, the EE begins to drop, an indication of an order that is being built, which is the polarized phase. 

\subsection{Three-leg ladder}
%\subsection{Ground state properties}
We start by examining the three lowest eigenstates of the three-leg ladder geometry. The energies of these states are portrayed in Fig.~\ref{fig:gapNy_3}. One can see that as system size increases, the gaps between these states decrease. 
\begin{figure*}
	\includegraphics[width=\textwidth]{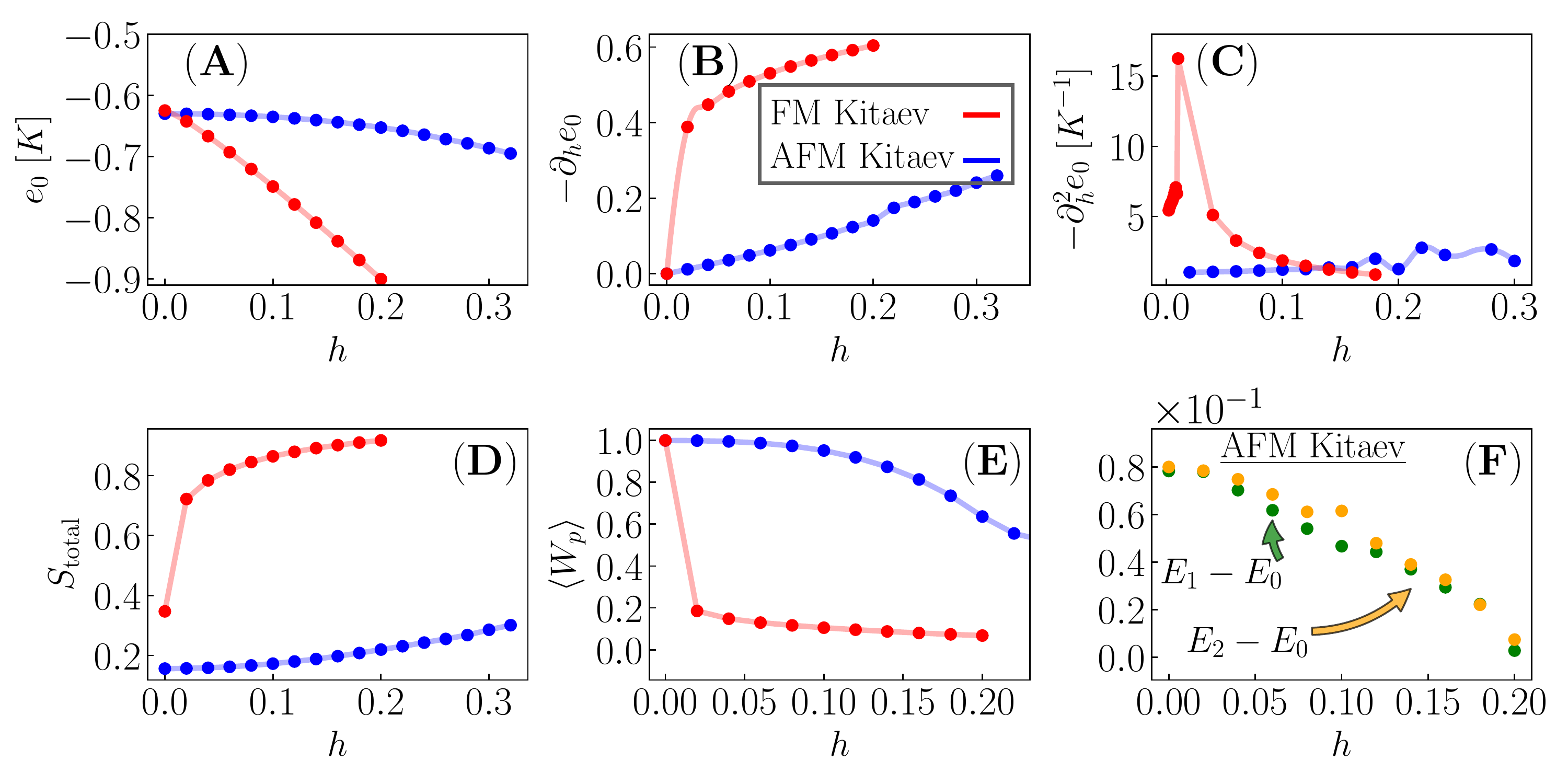}
	\caption{Magnetic phase diagram of the ground state of the $S=1$ ferromagnetic (red) and anti-ferromagnetic (blue) Kitaev model on a $24$-site three-leg ladder with cylindrical boundary conditions (appearing in Fig.~\ref{fig:Geom} A). The magnetic field, whose magnitude is $h$ (in units of $K$), is parallel to the $[111]$ direction. Panel A shows the ground state energy density, panel B is the magnetization density, panel C is the uniform magnetic susceptibility. Panel D and E are the total spin magnitude and the plaquette operator's, $W_p$, expectation value, respectively. Panel F shows the difference in energy between the first two excited states and the ground state. This quantity suggests that, similarly to the spin $S=1/2$ case, the specturm  shows a collapse at the phase transition to the intermediate phase. This transition is also captured by the uniform magnetic susceptibility.}
	\label{fig:MagPD_Ny_3}
\end{figure*}
We find that the ground state and the next two eigenstates are always in the $W_\ell=1$ sector. As shown in Fig.~\ref{fig:gapNy_3}, the energies of these three states become very close as the system size increases. Unfortunately, based on the finite size data obtained here, we cannot definitely determine whether the system has a degenerate ground state followed by an excitation gap or a gapless spectrum in the thermodynamic limit.
Similar to previous ED results~\cite{Koga_ED_2018} we find that as one increases the number of legs the ground state energy density actually increases, e.g., for the two-leg ladder we find $ e_0 \equiv \frac{E_0}{N} = -0.672~K$, and for the three-leg ladder $e_0 = -0.644~K$.

The magnetic field dependence of the ground state of the three-leg system is depicted in Fig.~\ref{fig:MagPD_Ny_3}. We plot the ground state energy density, the magnetization density, the uniform magnetic susceptibility, the total spin magnitude, and the plaquette operator expectation value (Fig.~\ref{fig:MagPD_Ny_3} A-E). We perform this analysis for both the FM and AFM interaction couplings. For the FM case, the field undermines the Kitaev interaction and easily polarizes the system (as can be seen in panels B-D). By examining the susceptibility on panel C, we conclude that the transition to the polarized phase occurs at $h_c=0.01$. However, for the AFM coupling, the field has to compete with the staggering nature induced by the Kitaev interaction. Therefore, the magnetization builds up much slower, and the total spin is increasing slowly as a function of the applied field (panel D). For the AFM case, panel F shows that the first two excited states energies decrease towards the ground state energy, which might indicate a spectrum collapse (similar to the two-leg ladder; see Fig.~\ref{fig:ED_18} for ED spectrum).    

The transition to the intermediate phase is not clear in the susceptibility, therefore we compute the EE and ES to investigate this further. The EE for the three-leg ladder as a function is summarized in Fig.~\ref{fig:EE}. Unlike for the two-leg ladder system, there is no jump in the ES at the Kitaev limit ($h=0$), implying there are no symmetry protected features in the EE unlike the two-leg geometry. The latter claim is also supported by the ES in Fig.~\ref{fig:ES_Ny_3}, which shows no degeneracy. In addition, as seen in Fig.~\ref{fig:EE}, the EE is monotonically building up towards the transition to the intermediate phase at $h=0.22$, beyond this field there are serious convergence issues.
 
We expect that the existence of the field-induced intermediate phase is robust and will occur in systems with larger circumference. However, based on studies of smaller clusters, the critical field values are expected to be affected by finite size effects. 

\section{Discussion}
In the current work we used DMRG and ED to investigate the nature of the ground states for the $S$=1 Kitaev model with both FM and AFM exchange interactions.
We presented these results via systematically studying the evolution of the ground states as a function of
the circumference size $L_y$ of the cylinder, and applied magnetic field.  Even though the DMRG method does not directly deal with the two-dimensional system, we use finite size scaling and topological properties of the ground state wave function, e.g. $W_\ell$, to infer properties of the two-dimensional limit.

Most notably, for our finite-size cylindrical clusters up of to $L_y = 6$ we found a number of numerical evidences , i.e. the even-odd effect of the Wilson loop operators $W_\ell$, the plaquette operators $W_p$, and lack of magnetic ordering, all suggesting that the ground states of the $S=1$ Kitaev model are quantum spin liquids.

First, we have identified an SPT for the $S=1$ two-leg ladder geometry via the two-fold degeneracy of the ground state and the degeneracy of the ES. As one introduces magnetic field for the AFM model, three phases are found: a highly-entangled disordered phase at weak fields, a gapless intermediate phase and a polarized phase. For the FM couplings, a direct transition to the polarized phase is found at a weaker field.
We find a similar magnetic phase diagram for the three-leg ladder.

We determine an upper bound on the excitation energy of the AFM model on a three-leg ladder cylinder -  $\Delta = 4 \times 10^{-2} \ K$. While this may suggest a quantum spin liquid with a small excitation gap, we cannot exlude the possibility of a gapless spin liquid  state in the thermodynamic limit. For example, if one linearly extrapolates these reported gaps to the thermodynamic limit one finds a  gapped spin liquid. On the other hand, it is not clear whether such a linear extrapolation is justified as the DMRG algorithm is not necessarily bound to find the actual first excited state. This makes us conclude that, given our data, we cannot distinguish between a gapless spin liquid or an existence of a small gap. Instead, we provide a numerical upper bound on the excitation gap.

We find great similarity between the $S=1/2$ and $S=1$ models:
(i) The ground-state is two-fold degenerate for the two-leg ladder. (ii) Both models share a similar response to applied magnetic field. In the AFM case, we obtained a phase diagram which is separated into three distinct regions (although the critical fields values differ by a factor of $\sim 1.5$ between $S=1/2$ and $S=1$ models). The critical field strengths marking the phase transition are accompanied by a spectrum collapse, which suggests a gapless disordered state.  (iii) The ground state energies of the FM and AFM couplings are the same. (iv) Both models exhibit extremely short-ranged correlations. The correlations become longer-ranged with applied field, together with a gradual drop in the magnitude of the (local) plaquette operators. 

Still, there are differences between the two models.  With no magnetic field, and contrary to the gapless spin liquid ground state of the $S$=1/2, this model shows a gapped ground state. Additionally, The even-odd effect we find for the $S=1$ model with respect to the ground state's $W_\ell$ sectors does not occur for the spin $S=1/2$ model.  
Another difference lies in the ES structure, while the $S=1/2$ two-leg ladder has four-fold degeneracy in its ES, the $S=1$ ES has a $2-4-2$ structure~\footnote{Interestingly, with applied field the ratio of the entanglement entropy between the two problems seems to be ${\rm EE}_{S=1}/{\rm EE}_{S=1/2}=\ln(3)/\ln(2)$, independent of $h$.}. This could be due to a different symmetry protecting the SPT.
Furthermore, note that for the soluble $S=1/2$ case the eigenvalues of the Wilson operators correspond to $\mathbb{Z}_2$ fluxes, and the ground state is in the $W_\ell= \pm 1$ sector depending on the boundary conditions (periodic versus anti-periodic boundary conditions for the Majorana fermions~\cite{Gohlke_2018}).The same flux sectors are present for the $S=1$ model, as was shown in this paper, however, the nature of the elementary excitations remain unresolved, which is an excellent question for future study.

Using the intuition previously obtained from spin $S=1/2$ calculations, we would like to point out that the two-leg ladder results may be used as a guide at a qualitative level to the two-dimensional limit. This is also seen by comparing the three-leg ladder phase diagram with that of the two-leg ladder in the presence of a magnetic field.

For the multi-leg ladders, the even-odd effect of the ground state's $W_\ell$ subspace , which we speculate to persist beyond $L_y=6$, suggests that in the thermodynamic limit, both $W_\ell= \pm 1$ sectors may be degenerate (since the excitation gap decreases as the system size is increased), hence it would restore a two-fold degeneracy of the ground-state on a cylinder. 

Based upon all the physical quantities we discussed 
above it is highly suggestive that the ground state of $S=1$ Kitaev model is a quantum spin liquid with a $\mathbb{Z}_2$ gauge structure. 

An open question remains whether a finite excitation gap exists in the presence of a magnetic field. As discussed earlier, whether the excitation gap in the thermodynamic limit, at zero and finite magnetic fields  (see Fig.~\ref{fig:hGap}), is finite or not is not completely resolved in this present work. However, by viewing the spectrum in the presence of a field (Fig.~\ref{fig:MagPD_Ny_3} F), one could conclude that the finite size gap is decreasing even further. If this is correct, it could mean that in the thermodynamic limit, the entire Kitaev phase may be gapless. However, a different option is that similarly to the $S=1/2$ model, a gap opens when the magnetic field is introduced, as it enters the chiral spin liquid phase. Then the system becomes gapless in the intermediate field regime, before it polarizes at sufficiently strong field.  Our current data cannot distinguish between these two scenarios. 

Note added: during the completion of this manuscript, we became aware of a similar analysis~\cite{Dong_2019} on the spin $S=1$ Kitaev model. We thank Donna N. Sheng for the correspondence and discussion about theirs and our results.
After completion of our manuscript, we became aware of two papers about the magnetic field affect on the $S=1$ Kitaev model~\cite{Zhu2020,Hickey2020}. Their main conclusions are consistent with ours.

\begin{acknowledgments}
This work was supported by the NSERC of Canada and the Center for Quantum Materials at the University of Toronto. YBK is further supported by the Killam Research Fellowship from the Canada Council for the Arts. HYK and PPS acknowledge support from the NSERC Discovery Grant 06089-2016. HYK acknowledges funding from the Canada Research Chair program.
We thank Nazim Boudjada, Matthias Gohlke, Jacob S. Gordon, Ciaran Hickey, and Hyunyong Lee for useful discussions.
\end{acknowledgments}

\newpage
%\section*{Supplementary Information}
\appendix
%\beginsupplement
\section{Finite size excitation gap in the presence of a weak magnetic field}
\begin{figure}
	\includegraphics[width=\linewidth]{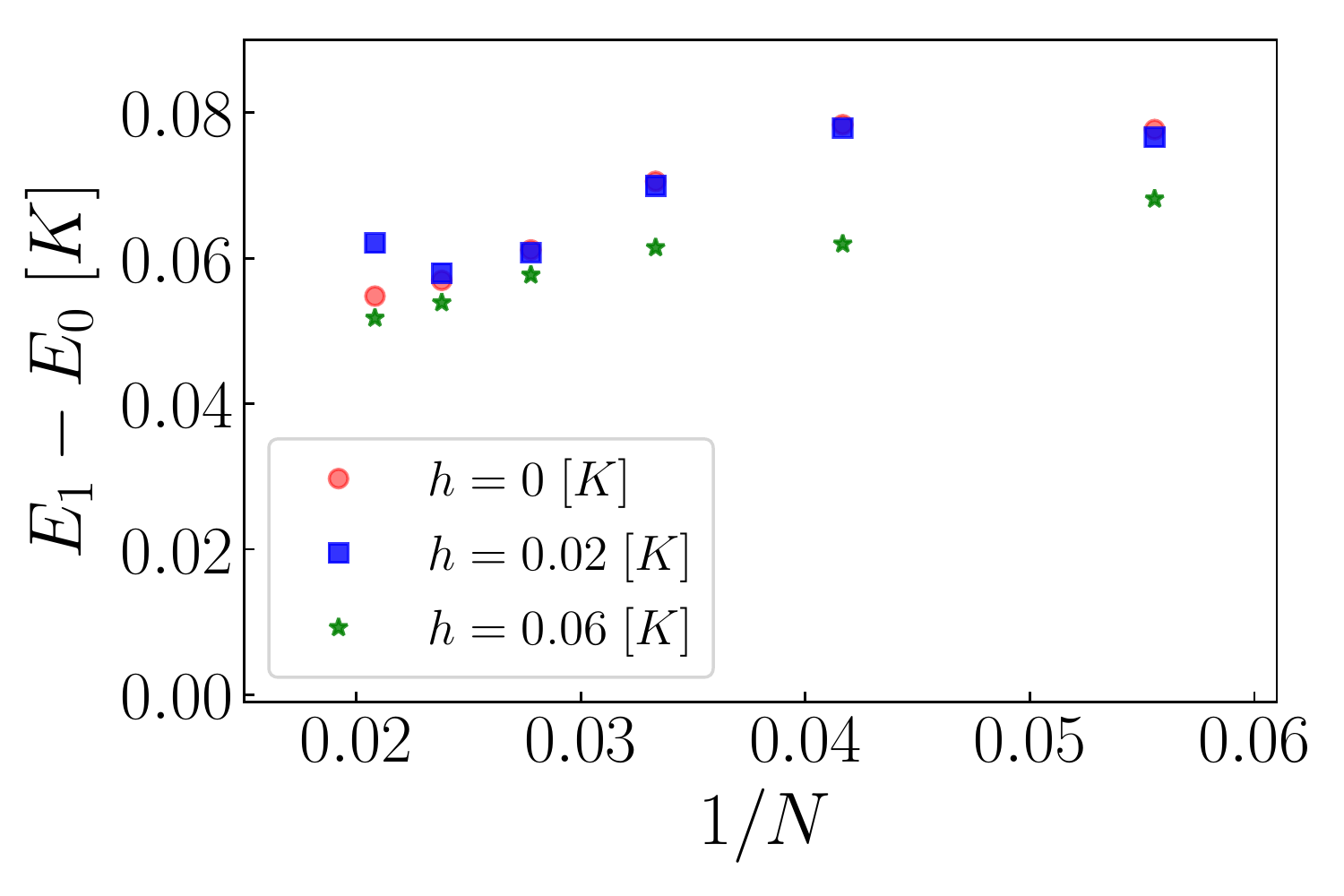}
	\caption{Ground state and first excited state energy difference, $E_1-E_0$, as a function of system size, $N$, for the three-leg ladder, $L_y=3$, geometry. Red circles denote the Kitaev limit, as in Fig.~2 (main text), blue square represent $h=0.02$ and green stars are $h=0.06$. While the energy density difference seems to decrease with system size, it is hard to conclude whether the gap remains in the presence of a magnetic field in the thermodynamic limit.}
	\label{fig:hGap}
\end{figure}

Similarly to the discussion for $h=0$ in the main text, we show the ground state and the first excited state energy densities for three-leg ladder clusters with up to $N=54$ sites in Fig.~\ref{fig:hGap}. In the presence of weak magnetic fields ($h=0.02,~0.06$), we see that as one increases the system size the gap, $E_1-E_0$ between the first excited state and the ground state decreases, however we cannot distinguish whether in the thermodynamic limit a small finite gap remains or the system is gapless.

\section{Spin-half magnetic phase diagram}
The magnetic field phase diagram of the $S=1/2$ Kitaev model has been studied extensively~\cite{Zhu_2018,Gohlke_2018,Hickey_2019,Patel_2019}. Focusing on the AFM model, it was found that finite field strength closes the vison gap and a gapless $U(1)$ spin liquid emerges. Upon further increase of the magnetic field, there is a second transition to the high-field (partially) polarized paramagnet. These transitions are captured by the $[111]$-magnetization, which shows two kinks corresponding to the chiral spin liquid to $U(1)$ spin liquid transition, and at a higher field a transtion to the polarized state. As a consequence of the kinks in the magnetization, the magnetic susceptibility shows a two-peak structure. By examining ED studies~\cite{Zhu_2018,Hickey_2019} on small clusters, one sees that near the two critical fields, the spectrum collapses.

The plaquette operator $W_p$, is featureless as a function of the applied magnetic field, and does not show any signature for the transition. For non-zero field, $\vec{h}$, it is no longer a conserved quantity, and as one increases the field strength it decreases in value. The plaquette operator is $W_p \approx 1$ in the Kitaev spin liquid phase and $W_p \approx 0$ in the polarized phase. As for the intermediate phase, it simply interpolates between these two limiting values.
This is one indication that the plaquette flux strongly fluctuates in the intermediate phase, which is consistent with a gapless disordered state, and was identified~\cite{Hickey_2019} as a gapless spin liquid (GSL).

\section{Entanglement spectrum}
\begin{figure}
	\includegraphics[width=\linewidth]{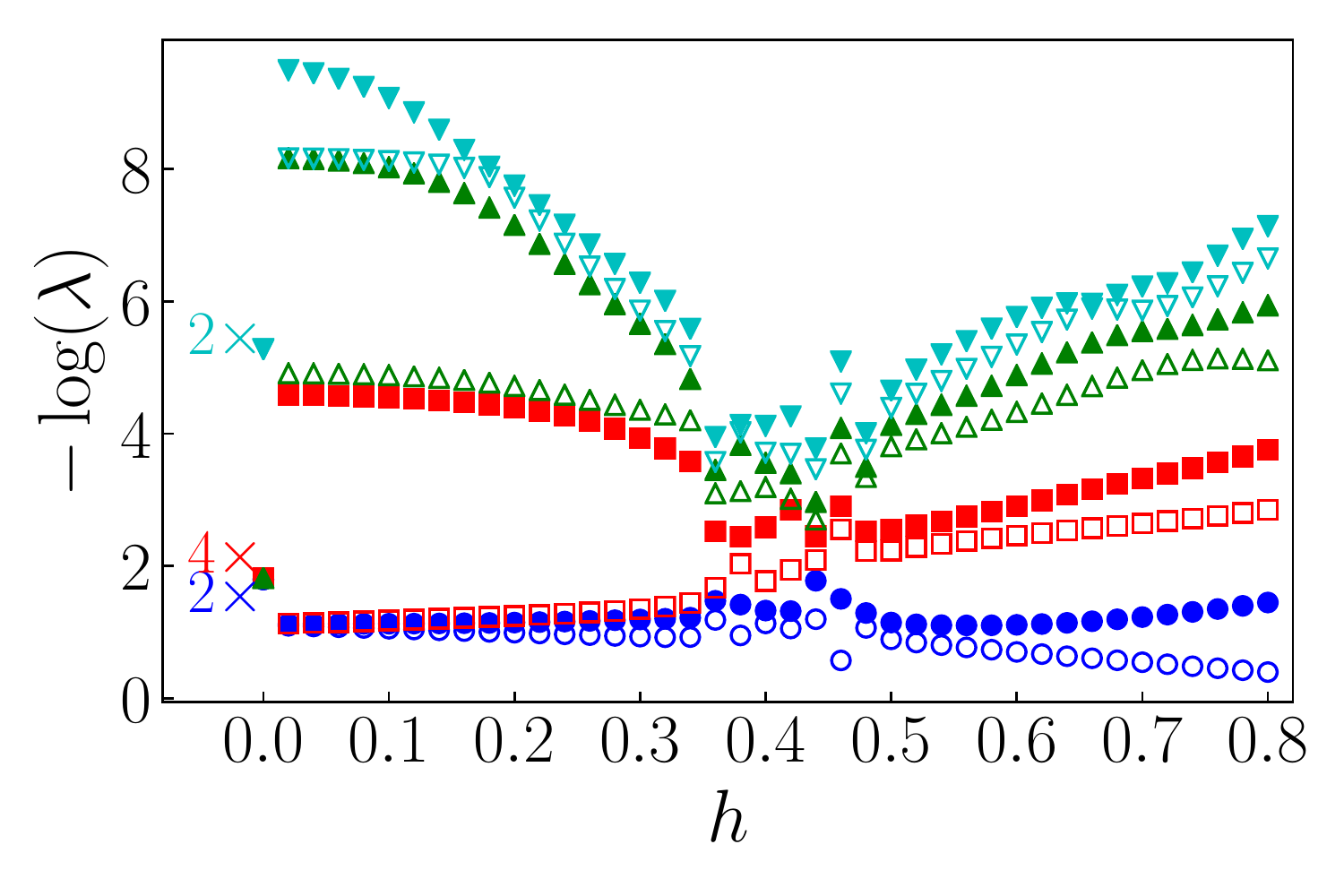}
	\caption{Entanglement spectrum (ES) partitioned with a cut on the middle rung of the two-leg, $L_y=2$, system as a function of magnetic field $h$. For convenience we show here only the $8$ largest Schmidt coefficients. At $h=0$, the ES has a $2-4-2$ degeneracy structure, this degeneracy is depicted by the numbers to the left of the markers. As can be seen in this figure, the degeneracy is broken as finite magnetic field is introduced.}
	\label{fig:ES_Ny_2}
\end{figure}

\begin{figure}
	\includegraphics[width=\linewidth]{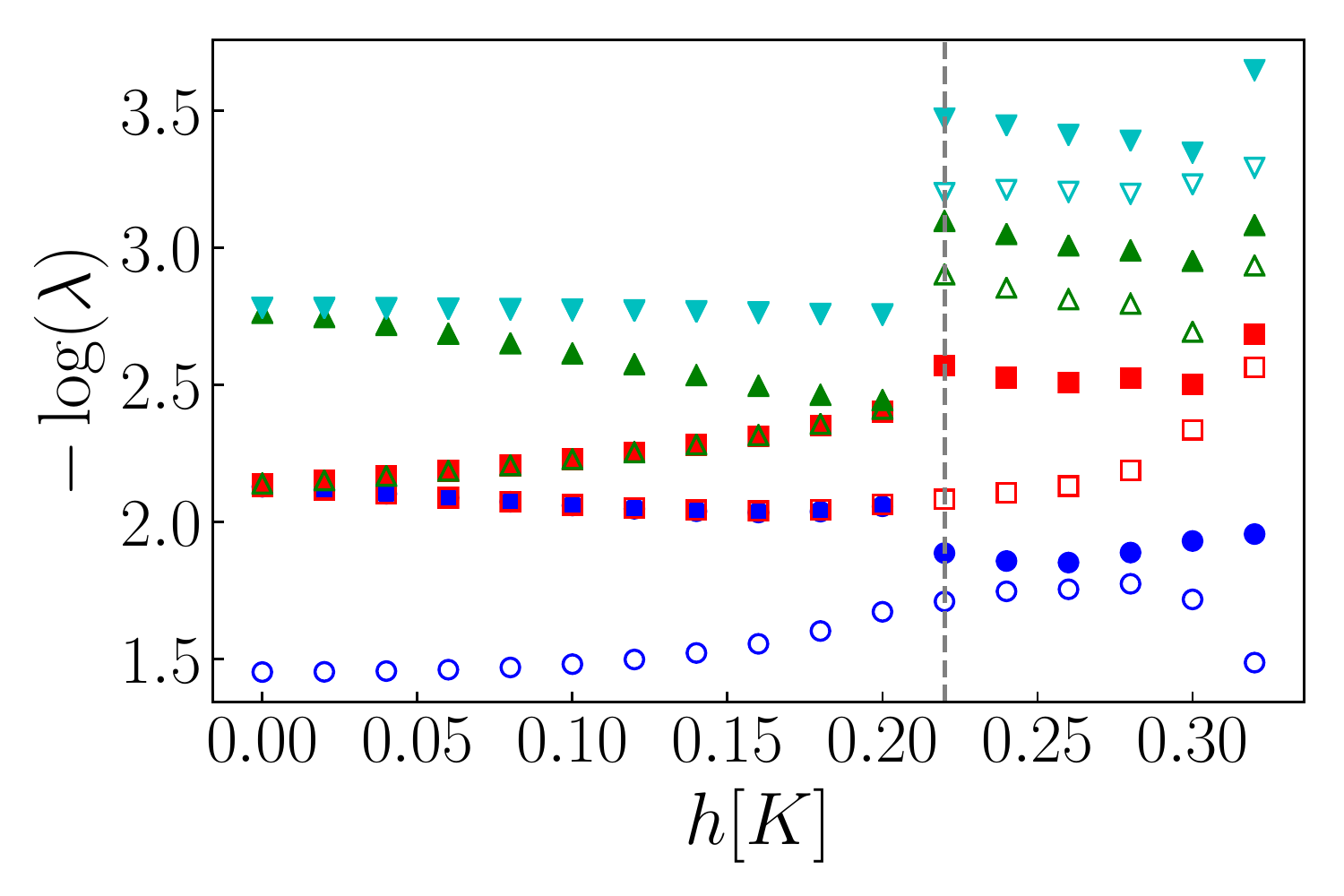}
	\caption{Entanglement spectrum (ES) partitioned with a cut on the middle rung of the three-leg, $L_y=3$, system as a function of magnetic field $h$. For convenience we show here only the $8$ largest Schmidt coefficients. The ES does not show any degeneracy structure. However, a transition can be seen as a discontinuity of the Schmidt eigenvalues occurring near $h=0.22$, which is shown by the dashed line.}
	\label{fig:ES_Ny_3}
\end{figure}

The two-leg geometry and three-leg geometry differ substantially at zero field. The first, being an SPT, is characterized by the degenerate pattern of its Schmidt eigenvalues as seen in Fig.~\ref{fig:ES_Ny_2}. The three-leg lacks this degeneracy, as seen in Fig.~\ref{fig:ES_Ny_3}, and the transition to the intermediate phase, as described in the main text, is found around $h=0.22$, where a discontinuity of the ES occurs. 

\section{Bilayer Kitaev model}
One can gain some intuition by studying a $S$=1/2 bilayer model, which at the appropriate limit would mimic its spin-one counterpart,
\be
\mathcal{H}_{\rm BL} = K \sum_{\substack{\gamma \\ \left\langle i,j \right\rangle_\gamma}} \sum_{n=1}^2 S_{i,n}^\gamma S_{j,n}^\gamma - \vec{h} \cdot \sum_{i} \sum_{n=1}^2 \vec{S}_{i,n} - J\sum_i \vec{S}_{i,1} \cdot \vec{S}_{i,2}, 
\label{eq:Ham2}\ee
where $S_{i,n}^\gamma = \frac{1}{2}\sigma_{i,n}^\gamma$, and $\sigma_{i,n}^\gamma$ is a Pauli matrix at the $i$-th site and layer $n=1,2$. $J$ is the ferromagnetic Heisenberg interaction between two spins on different layers. This bilayer model was shown to be useful for calculating thermodynamic properties of the $S=1$ Kitaev model~\cite{Seifert_2018,Koga_2018,Tomishige_2018,Tomishige_2019}. These ED and TPQ studies show the characteristic two-peak structure of the specific heat for strong $J$, the excitation energy, and the overall phase diagram as a function of the inter-layer coupling $J$.
\begin{figure}
	\includegraphics[width=\linewidth]{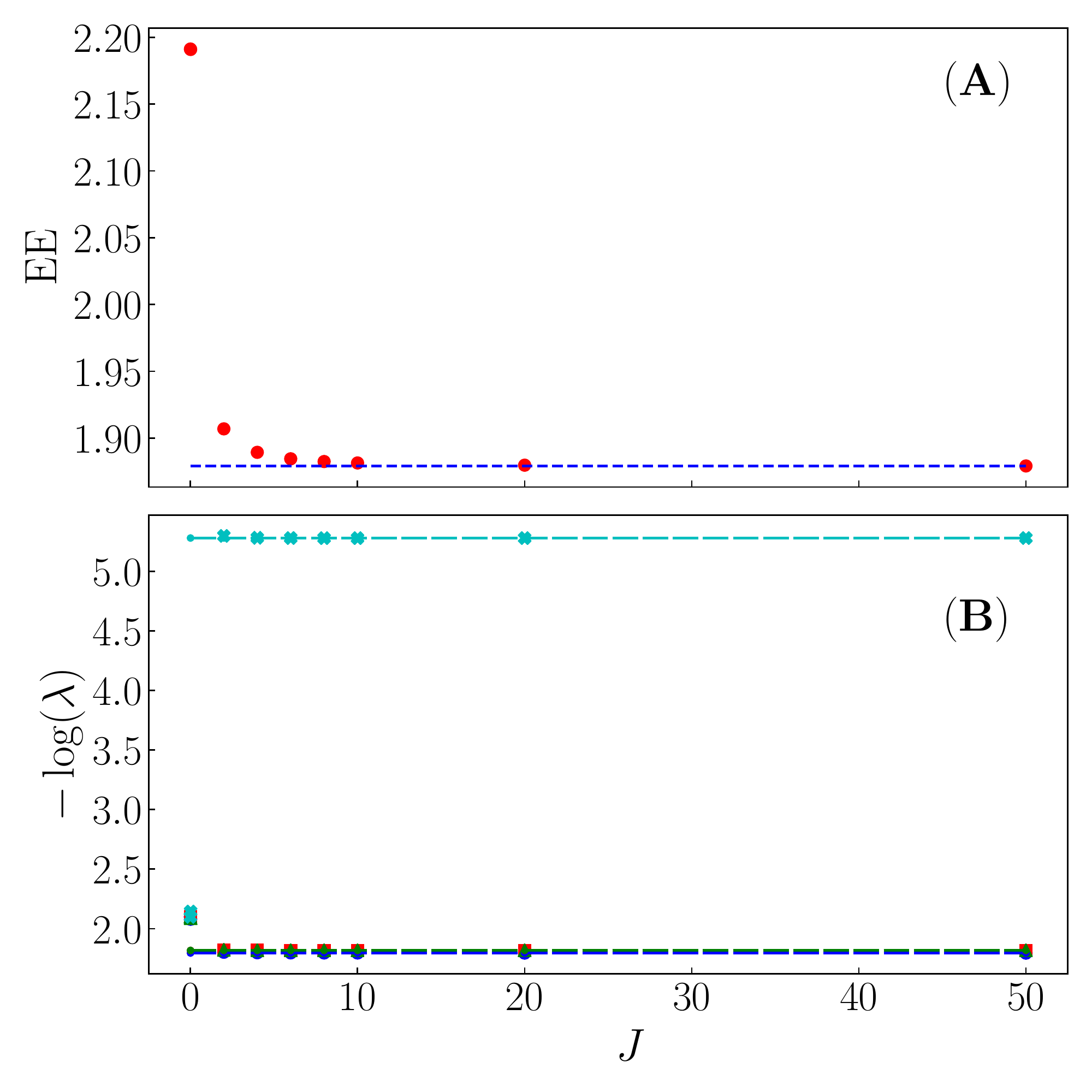}
	\caption{(A) Entanglement entropy (EE) and, (B) entanglement spectrum (ES), are shown for the $S$=1/2 bilayer model as a  function of ferromagnetic inter-layer coupling $J$. The system is partitioned with a cut on the middle rung of the two-layer.  The values approach those found in the $S$=1 Kitaev model with $h=0$ at finite $J/K \lesssim 10$, shown in Fig.~4 (main text) and Fig.~\ref{fig:ES_Ny_2} (denoted here as dashed lines). Note that the degeneracy of the Schmidt eigenvalues is also restored at this limit of $J/K$ (2-4-2 degeneracy structure).}
	\label{fig:2cpld}
\end{figure}
Here, we would like to target the $S=1$ limit of this model. We argue that the phase transition from two decoupled (or weakly coupled) layers occurs at finite values of $J$.
As can be seen in Fig.~\ref{fig:2cpld}, the EE and ES of the coupled-$S=1/2$ bilayer two-leg ladder as a function of the ferromagnetic inter-layer coupling, $J$. We converge to values obtained for the pure $S=1$ Hamiltonian, shown in Fig.~4 (main text) and Fig.~\ref{fig:ES_Ny_2}. This suggests that a phase transition from two decoupled, or even weakly coupled $S=1/2$ chains to a Kitaev $S$=1 system occurs at finite coupling strength $J$.

\section{Exact diagonalization results}
\begin{figure}[h!]
	\includegraphics[width=\linewidth]{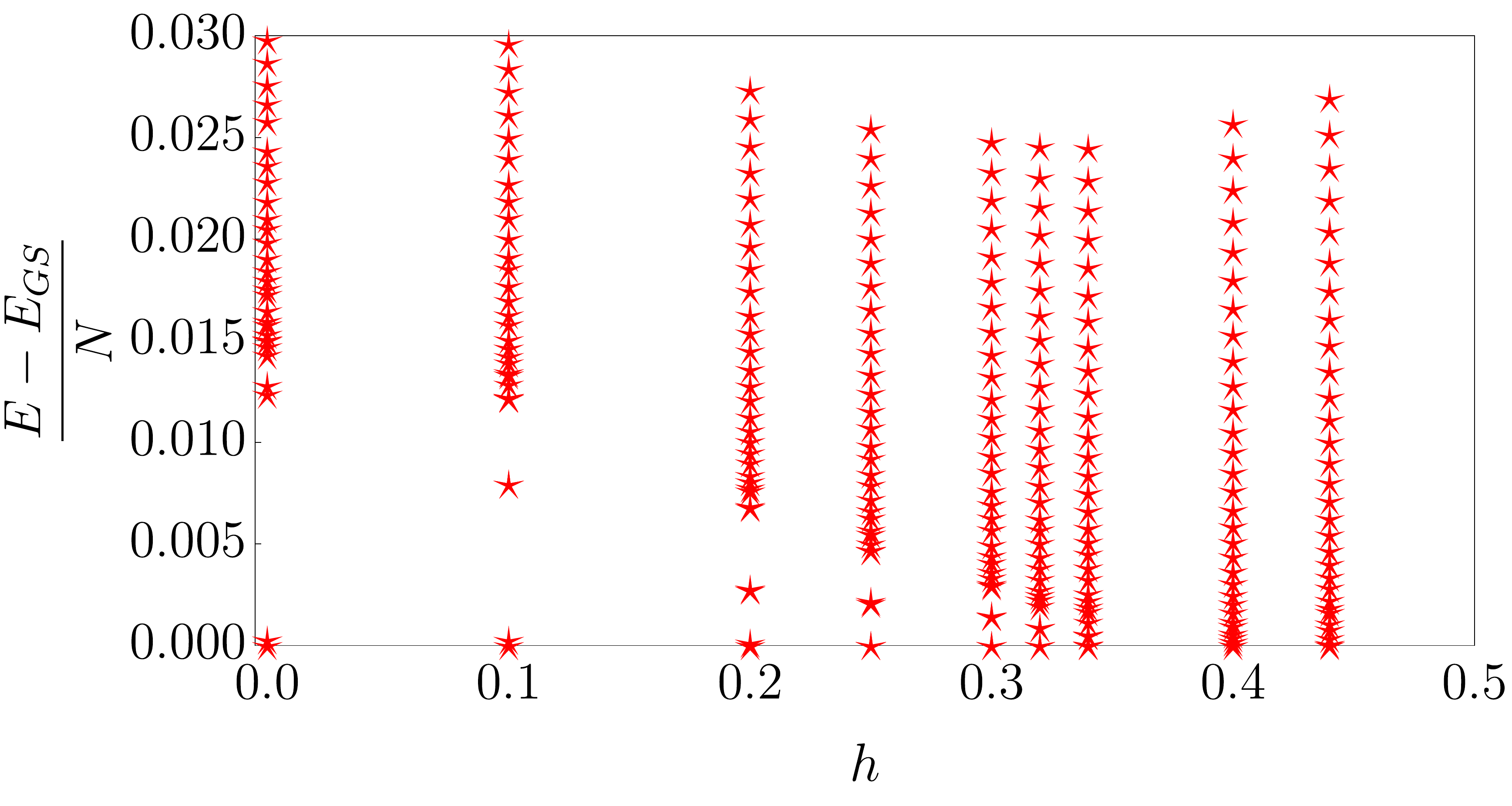}
	\caption{ED spectrum for $18$-site two-leg ladder, as a function of uniform magnetic field $h$. A spectrum collapse can be seen in the vicinity of $h=0.3~K$.}
	\label{fig:ED_18}
\end{figure}

\begin{figure}[h!]
	\includegraphics[width=\linewidth]{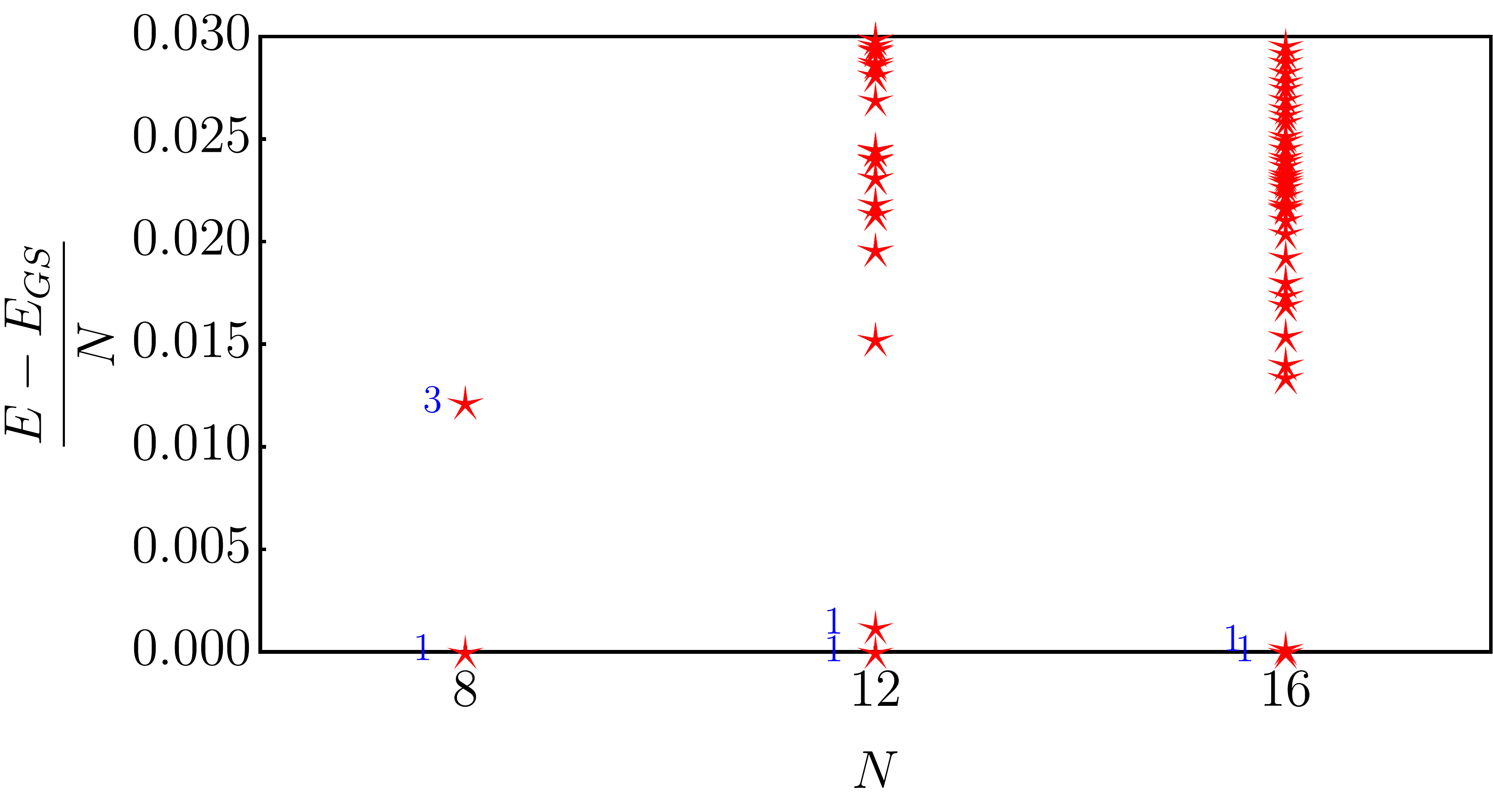}
	\caption{ED spectrum for various two-leg ladder with sizes $N$ and periodic boundary conditions (PBC). The degeneracy of the lowest eigenstates is marked by the numbers next to the markers. The two-fold degeneracy of the ground state is seen starting from $N \geq 12$.}
	\label{fig:ED_BC}
\end{figure}

We have performed ED on the AFM model on up to $18$-sites two-leg ladder. 
In Fig.~\ref{fig:ED_18} we show the spectrum collapse occurring in the vicinity of $h=0.3 ~K$. In addition, in Fig~\ref{fig:ED_BC} we show the finite size gap and the degeneracy of eigenstates for various system sizes with periodic boundary conditions. Note that the two-fold degeneracy occurs for system sizes $N \ge 12$. 


\section{Additional symmetries of the two-leg ladder}
\begin{figure}[b!]
	\includegraphics[width=\linewidth]{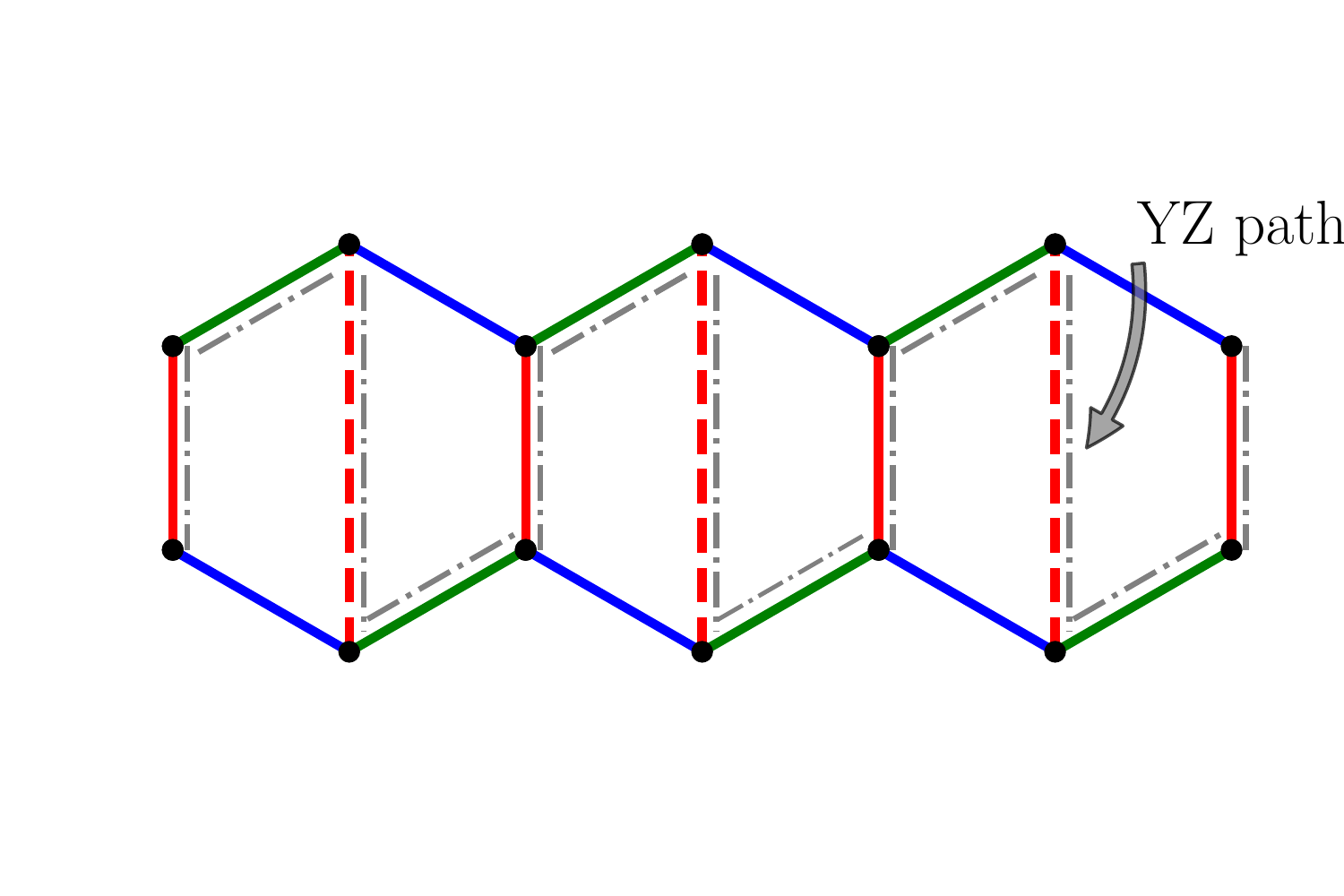}
	\caption{Two-leg ladder ($L_y=2$) geometry, which is equivalent to a square ladder. ``YZ path" is used to define the $\mathcal{O}_x$ string operator in Eq.~\ref{eq:Wx}. }
	\label{fig:Geom2}	
\end{figure}

Besides the plaquette operator $W_p$, the two-leg ladder has two more constants of motion (since $W_\ell$ defined in Eq.~3 (main text) is invalid for this geometry)
\be
\mathcal{O}_x = \prod_{ i \in \text{YZ path}} e^{i\pi S_i^x},
\label{eq:Wx}\ee
and similarly    
\be
\mathcal{O}_y = \prod_{ i \in  \text{XZ path}} e^{i\pi S_i^y},
\label{eq:Wy}\ee
where XZ-path runs along $x$ and $z$-bonds (and similarly for YZ-path), as depicted in Fig.~\ref{fig:Geom2}.
For the two-leg ladder, the degenerate ground states are distinguishable by $\mathcal{O}_x$, $\mathcal{O}_y$, and $W_z$. In a finite system (where finite gap still exists between these two states), all the operators mentioned are $+1$ for the lowest eigenstate, and $-1$ for the next eigenstate.

It is interesting to note another commuting operator, which is the open string operator along $x$-direction (the axis of the cylinder) 
\be
W_z = \prod_{ i \in \rm x-string} e^{i\pi S_i^z},
\ee
where $x$-string is an open string along the $x$-direction, which includes every spin from one end of the finite cylinder to the other end.
In a torus geometry (periodic boundary conditions along $x$-direction),
$W_z$ will become the Wilson loop operator for a closed loop along the $x$-direction. However, its significance for the cylinder could be further studied. 
We find that for the three-leg geometry the ground state is always in the $W_\ell=1$ sector and $W_z=1$. For all the cluster sizes we examined, the first excited state also has $W_\ell=1$ and $W_z=1$. The next excited state, however, shows $W_\ell=1,\enspace W_z=-1$.

\end{document}